\documentclass[aps,twocolumn,preprintnumbers,nofootinbib,superscriptaddress,10pt]{revtex4-1}

\usepackage{amsmath}
\usepackage{graphicx}
\usepackage{times}
\usepackage{amssymb}
\usepackage[colorlinks]{hyperref}
\usepackage{textcomp}
\usepackage{xcolor}
\usepackage[T1]{fontenc}
\usepackage{ulem}

\begin{document}
\title{Electromagnetically induced transparency with Cu$_2$O excitons in the presence of phonon coupling}
\author{V. Walther$^\dagger$}
\email[Electronic address: ]{valentin.walther@cfa.harvard.edu}
\affiliation{Aarhus Universitet, Institut for Fysik og Astronomi, Ny Munkegade 120, 8000 Aarhus C, Denmark}
\affiliation{ITAMP, Harvard-Smithsonian Center for Astrophysics, Cambridge, Massachusetts, USA}
\author{P. Gr\"unwald$^\dagger$}
\affiliation{Aarhus Universitet, Institut for Fysik og Astronomi, Ny Munkegade 120, 8000 Aarhus C, Denmark}
\author{T. Pohl}
\affiliation{Aarhus Universitet, Institut for Fysik og Astronomi, Ny Munkegade 120, 8000 Aarhus C, Denmark}

\begin{abstract}
Highly excited Rydberg states of excitons in Cu$_2$O semiconductors provide a promising approach to explore and control strong particle interactions in a solid-state environment. A major obstacle has been the substantial absorption background that stems from exciton-phonon coupling and lies under the Rydberg excitation spectrum, weakening the effects of exciton interactions. Here, we demonstrate that two-photon excitation of Rydberg excitons under conditions of electromagnetically induced transparency (EIT) can be used to control this background. Based on a microscopic theory that describes the known single-photon absorption spectrum, we analyze the conditions under which two-photon EIT permits separating the optical Rydberg excitation from the phonon-induced absorption background, and even suppressing it entirely. Our findings thereby pave the way for the exploitation of Rydberg blockade with Cu$_2$O excitons in nonlinear optics and other applications.
\end{abstract}

\maketitle

\def\thefootnote{$\dagger$}\footnotetext{These authors contributed equally to this work}
The ability to control and manipulate highly excited Rydberg states of atoms has led to numerous breakthroughs, from simulating quantum magnetism~\cite{Schauss2010,Schauss2015,Labuhn2016,Zeiher2016,Zeiher2017,Bernien2017,Keesling2019,Leseleuc2019} and performing quantum gates~\cite{Levine2019,Graham2019,Madjarov2020}, to quantum and nonlinear optics~\cite{Raimond2001,dudin2012,peyronel2012,Li2013,gorniaczyk2016,thompson2017,Busche2017,Paris2017,murray2016,Firstenberg2016}.

In 2014, it was demonstrated that semiconductor excitons in bulk Cu$_2$O crystals can also be excited to Rydberg states which can be as large as their atomic counterparts~\cite{RydExc}. Their size equips them with remarkable properties in magnetic \cite{zielinska2017, schweiner2017, schweiner2017_2, schweiner2017_3, rommel2018, zieliska2019, assmann2016quantum} and electric \cite{zielinska2016, zielinska2018} fields, makes them very susceptible to free charges~\cite{heckoetter2018, heckoetter2020, walther2020plasma,Krueger2020}, suggests lasing applications \cite{ziemkiewicz2019} and permits integration into semiconductor microstructures \cite{krueger2018, steinhauer2020}. Importantly, the resulting long-range interactions of such large excitons~\cite{Valentin2018}, combined with the characteristic hydrogen-like scaling of the energy levels and lifetimes \cite{heckoetter2017}, can give rise to a strong suppression of Rydberg excitation, high optical nonlinearities~\cite{walther2018giant, khazali2017} and topological states \cite{poddubny2019topological}. A key prerequisite for exploiting such interaction effects in future experiments and potential applications is the availability of isolated Rydberg excitation lines. While this is excellently satisfied in atomic systems, the observed excitonic Rydberg series in Cu$_2$O rests on a broad background that accounts for more than 50\% of the optical density on some Rydberg-state resonance. Stemming from the simultaneous excitation of an optical phonon and a deeply bound exciton~\cite{elliott1957,Flo2017}, this background represents an additional excitation channel that inevitably competes with the Rydberg-exciton series. Since such a competition naturally weakens many effects of excited-state interactions, controlling the phonon-induced absorption background and resonant Rydberg-state excitation remains an important challenge.   

\begin{figure}[t]
\includegraphics[width=\columnwidth]{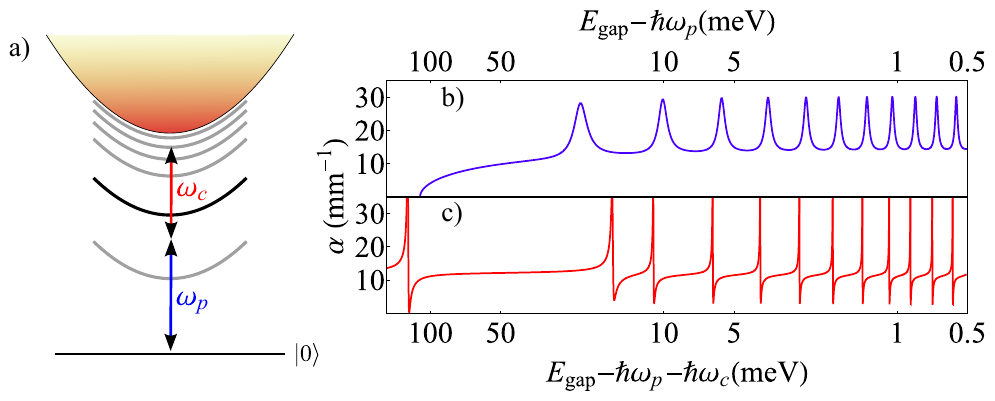}
\caption{EIT in Cu$_2$O. (a) A weak probe laser with a frequency $\omega_p$ generates $p$-state excitons of the yellow series.  The simultaneous coupling to a continuum of phonon-dressed exciton states gives rise to the typical absorption spectrum of Cu$_2$O below the band gap energy $E_\text{gap}$ of the yellow series, as shown in (b). As further illustrated in panel (a), an additional control laser field with a frequency $\omega_c$ couples the generated excitons to an excited $s$-state of the yellow series to establish conditions of electromagnetically induced transparency.  As shown in panel (c), this makes it possible to control the spectroscopic properties of the material and leads to a series of narrow absorption dips at a given $s$-state resonance with a greatly suppressed absorption background at each minimum.} \label{fig.system}
\end{figure}

In this work, we demonstrate that this can be achieved via electromagnetically induced transparency (EIT)~\cite{boller1991} involving excited exciton states. Two-photon excitation of long-lived excitons in $s$-states via a low-lying $p$-state~\cite{Artoni2000, Bassani2004} [see Fig.~\ref{fig.system}(a)], hereby leads to interference of the two excitation pathways that can be used to
eliminate the population of the strongly coupled and rapidly decaying intermediate state \cite{li2014}. Here, we show that this effect cannot only be employed to achieve narrow Rydberg-exciton lines but also allows to control and even reduce the broad phonon background that would otherwise supplement the exciton spectrum [see Fig.~\ref{fig.system}(b,c)]. Depending on the material parameters, we identify three distinct regimes in which either the discrete exciton lines or the broad phonon background are suppressed individually, or where both vanish simultaneously to yield transparency on two-photon resonance. The results of this work, thereby, suggest novel experimental probes of optical processes and phonon effects in semiconductors, and open the door to nonlinear optics and explorations of many-body phenomena with interacting Rydberg excitons.

\begin{figure}[t]
\includegraphics[width=\columnwidth]{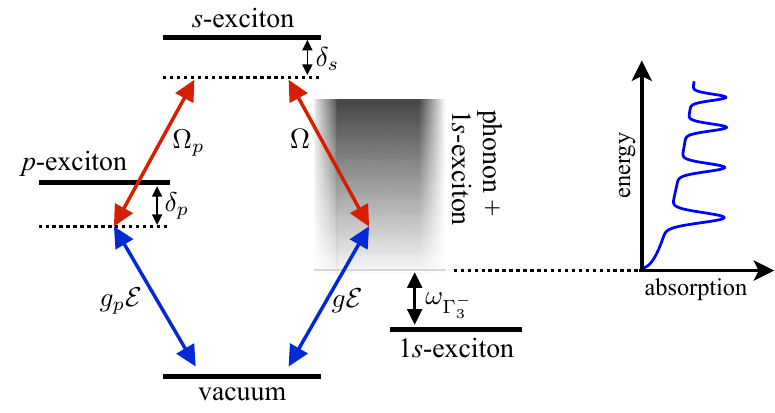}
\caption{The incident probe laser with electric field amplitude $\mathcal{E}$ excites $p$-state excitons with an optical coupling strength $g_p$ and detuning $\delta_p$. Simultaneously, the field couples to the continuum of phonon-dressed $1s$-excitons with coupling strength $g$. The dressed states start to appear at an energy $\hbar \omega_{\Gamma_3^{-}}$ above the $1s$-state. The combination of the $p$-state excitation and the phonon-dressed excitons generate the broad absorption spectrum sketched on the right [cf. Fig.~\ref{fig.system}(b)].
As we show in this work, this typical absorption can be controlled by a second light field that couples both types of excitons to an $s$-state with coupling strengths $\Omega_s$ and $\Omega$, respectively, and a two-photon detuning $\delta_s$.} \label{fig.setup}
\end{figure}

Most spectroscopic studies of Cu$_2$O are based on the excitation of $p$-state excitons of the so-called yellow series, with wavelengths of $\sim590$~nm \cite{schweiner2016, thewes2015, schoene2016,zielinska2016_2, konzelmann2019}. Around each resonance, the light field couples to a $p$-state exciton that can be described by a bosonic field operator $\hat{X}_p$, and is excited with a linewidth $\Gamma_p$, a coupling strength $g_p$ and a frequency detuning $\hbar \delta_p=E_p-\hbar \omega_p$, defined by the exciton energy $E_p$ and laser frequency $\omega_p$. However, early measurements of the series~\cite{RydExc} have established the presence of a broad absorption background that starts above the energy $E_0$ of the $1s$-ortho exciton and extends all the way across the band edge [Fig.~\ref{fig.system}(b)]. In the underlying absorption process, a photon virtually excites an intermediate state, which subsequently decays into an optical phonon and a $1s$-ortho exciton~\cite{elliott1957}. This indirect excitation process is necessary since neither the generation of the phonon nor the excitation of the $1s$ exciton is dipole-allowed. The dominantly involved states have recently been identified as the $s$-state excitons from the blue series and the optical phonons from the $\Gamma_3^-$-branch~\cite{Flo2017}. 
The underlying exciton-phonon scattering process can well be described by a Fr\"ohlich interaction term
\begin{equation}
	\begin{split}
		\hat H_\text F=&\sum\limits_{\mu}\int {\rm d}^3q\:{\rm d}^3q^{\prime}\,h_{\mu}(\mathbf q,\mathbf q^{\prime})\\
		&[\hat b^\dagger(-\mathbf q)\hat X_{1s}^\dagger(\mathbf q+\mathbf q^{\prime})\hat X_\mu(\mathbf q^{\prime})+\text{h.c.}],
	\end{split}
\end{equation}
where low-lying excitons from the blue series at momentum $\mathbf q'$, described by the operator by $\hat{X}_\mu (\mathbf q^{\prime})$, decay into a pair of a $1s$-exciton at momentum $\mathbf q+\mathbf q^{\prime}$, described by $\hat{X}_{1s}(\mathbf q+\mathbf q^{\prime})$, and a phonon that carries the excess momentum and is described by the bosonic operator $\hat{b}(-\mathbf q)$. The rate of this process is determined by the coupling constant $h_{\mu}(\mathbf q,\mathbf q^{\prime})$. While the kinetic energy of the $1s$-exciton of mass $m$ and rest energy $E_0$ leads to a momentum-dependent detuning $\hbar\delta_0(\mathbf q) = E_0+\hbar^2q^2/(2m) - \hbar \omega_p$, the phonon dispersion, $\hbar \omega_{\Gamma_{3}^-}$, can be taken as constant across the relevant phonon momenta. The blue exciton states are far off-resonant ($|\delta_\mu| \approx 0.4$ eV) and therefore only weakly populated. One can, thus, apply adiabatic elimination of their fast dynamics to describe the overall scattering process in terms of the composite bosons $\hat X(\mathbf q,\mathbf q')=\hat b(-\mathbf q)\hat{X}_{1s}(\mathbf q+\mathbf q')$ (see App.~\ref{app:dressed_states}), with a complex linewidth $\gamma(\mathbf q, \mathbf q^\prime) = \Gamma + i (\delta_0(\mathbf q + \mathbf{q}^\prime) + \omega_{\Gamma_{3}^-})$ and effective optical coupling elements
\begin{align}
 g(\mathbf q,\mathbf q^\prime)=\sum_\mu g_\mu / \delta_\mu \cdot h_\mu(\mathbf q,\mathbf q^\prime) \label{eq:g_q}. 
\end{align}
It is this optical coupling to the associated continuum of states that leads to the broad absorption background described above. 

As we will show below, however, EIT can be used to control the phonon coupling through the formation of an excitonic dark state. Establishing EIT requires a third state that is relatively stable and can be excited via a two-photon process (Fig.~\ref{fig.setup}). A natural choice is a Rydberg $s$-state, which we describe by the operator $\hat{X}_s$. Hereby, another laser field with a frequency $\omega_c$ couples the generated $p$-state excitons to the long-lived Rydberg $s$-state with Rabi frequency $\Omega_s$ and frequency detuning $\hbar \delta_s = E_s - \hbar \omega_p - \hbar \omega_c$. The resulting three-level system features a dark state of the form $\sim -g_p /\Omega_s \hat{X}_s^\dagger |0\rangle$.

In addition, however, the control field laser also couples to the continuum states since $\hat{X}_p$ and $\hat{X}$ share the same parity. The associated Rabi frequency can be obtained by adiabatically eliminating the dynamics of the blue intermediate states as before yielding $\Omega(\mathbf q,\mathbf q^\prime)=\sum_\mu \Omega_\mu/\delta_\mu \cdot h_\mu(\mathbf q,\mathbf q^\prime)$. 
Like the isolated Rydberg states, the background alone can also establish an EIT dark state that carries no contribution from the intermediate states and is of the form $\sim - D_{g\Omega}/D_{\Omega^2} \hat{X}_s^\dagger |0\rangle$ with 
\begin{equation} 
	D_{\Omega^2}=\int {\rm d}^3q\frac{\Omega^2(\mathbf q,\mathbf q')}{\gamma(\mathbf q,\mathbf q')},\ D_{g\Omega}=\int {\rm d}^3q\frac{g(\mathbf q,\mathbf q')\Omega(\mathbf q,\mathbf q')}{\gamma(\mathbf q,\mathbf q')}. 
\end{equation}

Taken separately, the two excitation pathways, shown in Fig.~\ref{fig.setup}, therefore promote the formation of simple dark states with vanishing absorption on two-photon resonance. Their simultaneous presence, however, leads to a competition between each pathway that can be described by the following set of Heisenberg equations for the exciton operators
\begin{align}
 \dot{\hat{X}}_p({\mathbf q})=&-\gamma_p \hat{X}_p({\mathbf q})-ig_p\mathcal E({\mathbf q})-i\Omega_s \hat{X}_s({\mathbf q})\label{mt1}\\
 \dot{\hat{X}}_s({\mathbf q})=&-\gamma_s \hat{X}_s({\mathbf q})-i\Omega_s \hat{X}_p({\mathbf q})\nonumber\\
& +i\int {\rm d}^3 q^\prime\Omega(\mathbf q^\prime,\mathbf q) \hat{X}(\mathbf q^\prime,\mathbf q) \label{mt2}\\
 \dot{\hat{X}}(\mathbf q,\mathbf q^\prime)=&-\gamma(\mathbf q, \mathbf q^\prime) \hat{X}(\mathbf q,\mathbf q^\prime)+ig(\mathbf q,\mathbf q^\prime)\mathcal E({\mathbf q}^\prime)\nonumber\\
 &+i \Omega(\mathbf q,\mathbf q^\prime) \hat{X}_s({\mathbf q}^\prime)\label{mt3},
\end{align}
where we have defined $\gamma_{p/s} = \Gamma_{p/s} + i \delta_{p/s}$, and $\mathcal E$ denotes the amplitude of the probe-light field. Note that a generalization to multiple exciton states is straightforward [see Fig.~\ref{fig.system}(b)], and only leads to small interference effects for overlapping resonances \cite{grunwald2015signatures}.

These equations are readily solved for the steady-state expectation values, which yield the susceptibility 
\begin{equation}\label{eq:chi}
\chi^{(1)}(\mathbf q)\mathcal{E}(\mathbf q) =-\frac{g_p}{cn}\langle\hat X_p (\mathbf q)\rangle+\int {\rm d}^3\, q^\prime \frac{g(\mathbf q^\prime, \mathbf q)}{cn} \langle\hat{X}(\mathbf q^\prime,\mathbf q)\rangle
\end{equation}
where $n\approx 2.74$ is the dielectric constant of Cu$_2$O. The first term on the right hand side describes the coherence of the discrete $p$-state exciton, while the second term captures the contribution of the phonon-induced absorption background. The inverse absorption length in the crystal is then given by $\alpha = {\rm Im}(\chi^{(1)})$. Explicitly, we find from Eqs.~(\ref{mt1})-(\ref{eq:chi}) (see App.~\ref{app:opt_res})
\begin{align}
i\chi^{(1)}=\ &\frac{g_p}{cn} \frac{\Omega_sD_{g\Omega}-g_p({\gamma_s+D_{\Omega^2}})}{\gamma_p({\gamma_s+D_{\Omega^2}}) +\Omega_s^2}\nonumber\\
&-\frac{D_{g^2}}{cn}+ \frac{D_{g\Omega}}{{cn}}\frac{D_{g\Omega}\gamma_p +g_p\Omega_s}{\gamma_p({\gamma_s+D_{\Omega^2}}) +\Omega_s^2}, \label{eq.fullresponse}
\end{align}
with the additional parameter 
\begin{equation}\label{eq:Dg2}
D_{g^2}=\int d^3 q^\prime \frac{g^2(\mathbf q^\prime, \mathbf q)}{\gamma(\mathbf q^\prime, \mathbf q^\prime)}.
\end{equation}

\begin{figure}[t!]
\includegraphics[width=\columnwidth]{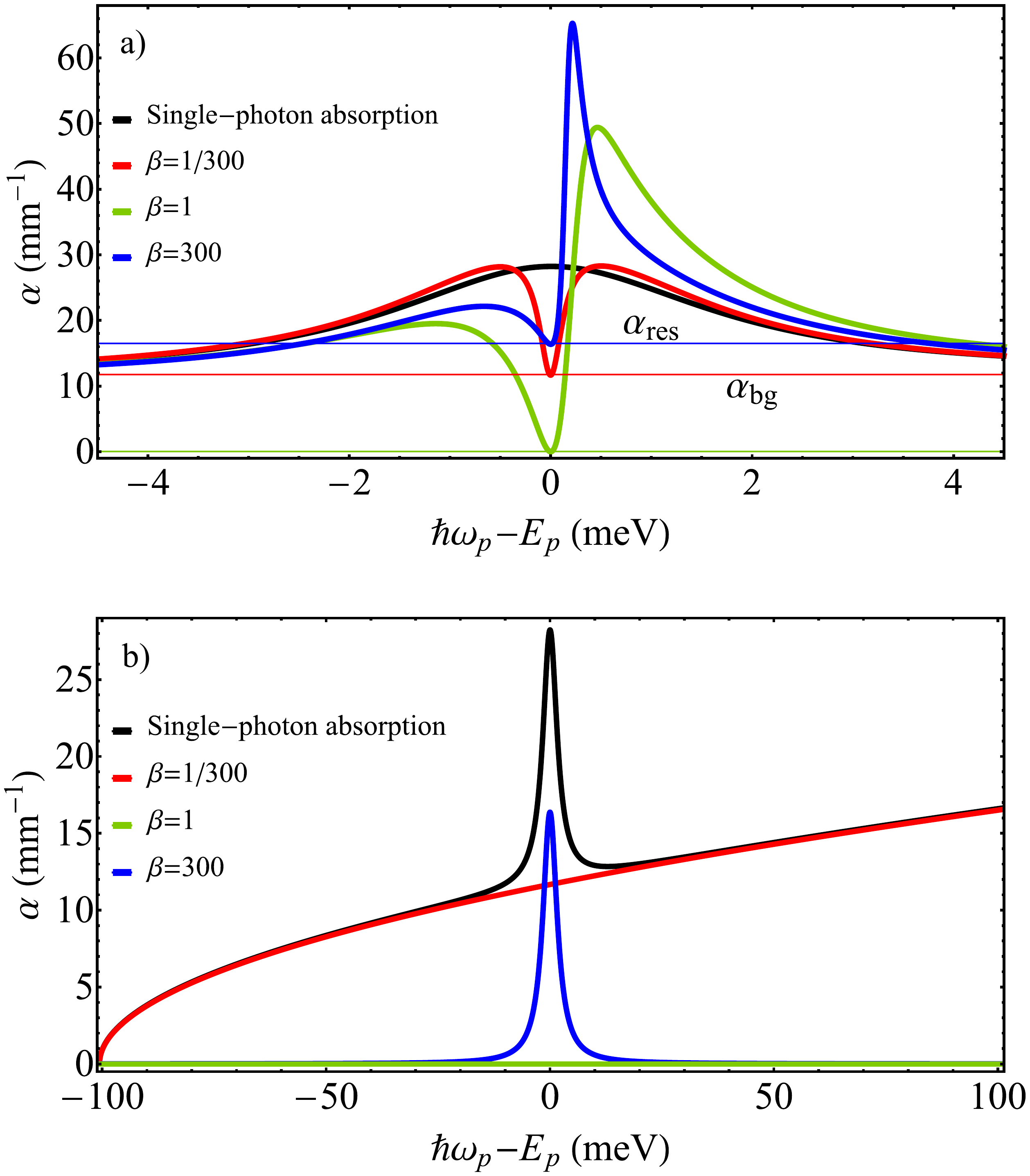}
\caption{Probe field absorption spectra around the $2p$-exciton at an energy $E_p$. In panel (a) the control field is tuned to resonance with the Rydberg-state transition, i.e. $\delta_p=\delta_s$. The parameters of the probe-field transitions are taken from~\cite{Flo2017}, and the Rydberg decay is neglected, $\Gamma_s=0$. The different colored curves show typical absorption profiles for different indicated values of $\beta = g_p\Omega/(g\Omega_s)$, whereby the larger of the two Rabi frequencies $\Omega$ and $\Omega_s$ is chosen to be $0.50$~meV. For comparison, the black line shows the absorption spectrum in the absence of EIT, i.e. for $\Omega_s=\Omega=0$. The thin horizontal lines indicate the resonant probe absorption stemming only from the $2p$-exciton ($\alpha_\text{res}$) and only from the background ($\alpha_\text{bg}$). Panel (b) shows the same quantities for identical parameters, but with the control field tuned to two-photon resonance, $\delta_s=0$.}\label{fig.EIT_ps}
\end{figure}

Without the coupling laser, the absorption is simply given by the sum of the yellow exciton and the background, $icn \chi^{(1)}=-g_p^2/\gamma_p - D_{g^2}$~\cite{Flo2017}. The observed characteristic square-root behavior of the latter \cite{baumeister1961} is recovered if we assume metastable continuum states ($\Gamma=0$, see App.~\ref{app:background}) and use a constant interaction coefficient over the range of relevant momenta, $h_\mu(\mathbf{q},\mathbf{q}^\prime) = \text{const}$. In this case, the two optical coupling elements are constant, $g(\mathbf{q},\mathbf{q}^\prime) = g$ and $\Omega(\mathbf{q}, \mathbf{q}^\prime) = \Omega$. 
Denoting the maximal kinetic energy of the $1s$ exciton as $\bar{E}$, Eq.~(\ref{eq:Dg2}) can be integrated straightforwardly across the entire Brillouin zone yielding
\begin{align}
	D_{g^2}=& \frac{8\pi^2g^2 m^{\tfrac{3}{2}}}{\sqrt{2}\hbar^2} \left( \sqrt{\hbar\omega_p-E_0-\hbar \omega_{\Gamma_3^-}} -i \frac{2}{\pi}\sqrt{\bar{E}}\right)\label{eq.Dg2fin}
\end{align}
for probe laser energies $\hbar \omega_p \geq E_{0}+\hbar \omega_{\Gamma_3^-}$. This simple expression indeed describes the experimentally observed $\omega_p$-dependence to a high accuracy \cite{baumeister1961}, justifying the simplification of the exciton phonon coupling, $h_\mu(\mathbf{q},\mathbf{q}^\prime) = \text{const}$. This also simplifies the parameters $D_{\Omega^2}=\frac{\Omega^2}{g^2}D_{g^2}$ and $D_{g\Omega}=\frac{\Omega}{g}D_{g^2}$ and permits reducing the total susceptibility to
\begin{align}
	i\chi^{(1)}=&-\frac{\gamma_s(g_p^2+D_{g^2}\gamma_p)+(g_p\Omega-g\Omega_s)^2\left(\frac{D_{g^2}}{g^2}\right)}{\gamma_p(\gamma_s+\frac{\Omega^2}{g^2}D_{g^2})+\Omega_{s}^2}.\label{eq.resp_simp}
\end{align}
Since the Rydberg-state linewidth $\Gamma_s$ is typically small, the resonant absorption is largely determined by the second term in the numerator, which is controlled by the ratio $\beta = g_p\Omega/(g\Omega_s)$.

Figure~\ref{fig.EIT_ps}(a) shows the absorption profile for different values of this parameter. If $\beta=1$, the yellow $p$-states and the phononic background feature identical dark states. Under such conditions of congruent EIT, the two excitation pathways do not perturb each other, such that one can suppress absorption on two-photon resonance. For different values $\beta\neq1$, the competition between the pathways leads to a finite EIT dip. In the limit of $\beta \ll 1$, excitation via the yellow series is dominant and therefore experiences EIT, allowing the resonant absorption to drop to the background value $\alpha_\text{bg}$. In the opposite limit of $\beta \gg 1$, this background absorption can be suppressed via EIT and the residual absorption is that of the $p$-resonance, $\alpha_\text{res}$. 

If the control laser is tuned to two-photon resonance, the quantum state stays in the approximate dark state when scanning the $p$-resonance [Fig.~\ref{fig.EIT_ps}(b)]. Under conditions of congruent EIT, the system then becomes transparent for all probe detunings. Dominant EIT on the $p$-state excitons ($\beta \ll 1$) leads to an isolated phonon background that extends across the entire yellow series. On the other hand, if the phononic background experiences EIT ($\beta \gg 1$), it is entirely suppressed and one can realize a series of sharp isolated $p$-state resonances, as indicated by the blue line in Fig.~\ref{fig.EIT_ps}(b). 

It is interesting to note that the presence of the coupling laser can even increase the probe field absorption [see Fig.~\ref{fig.EIT_ps}(a)].
This effect occurs on the blue-detuned side of the resonance and is due to interference of the excitation pathways, where the phonon continuum yields an additional effective decay of the $s$-state.

\begin{figure}[t]
	\includegraphics[width=\columnwidth]{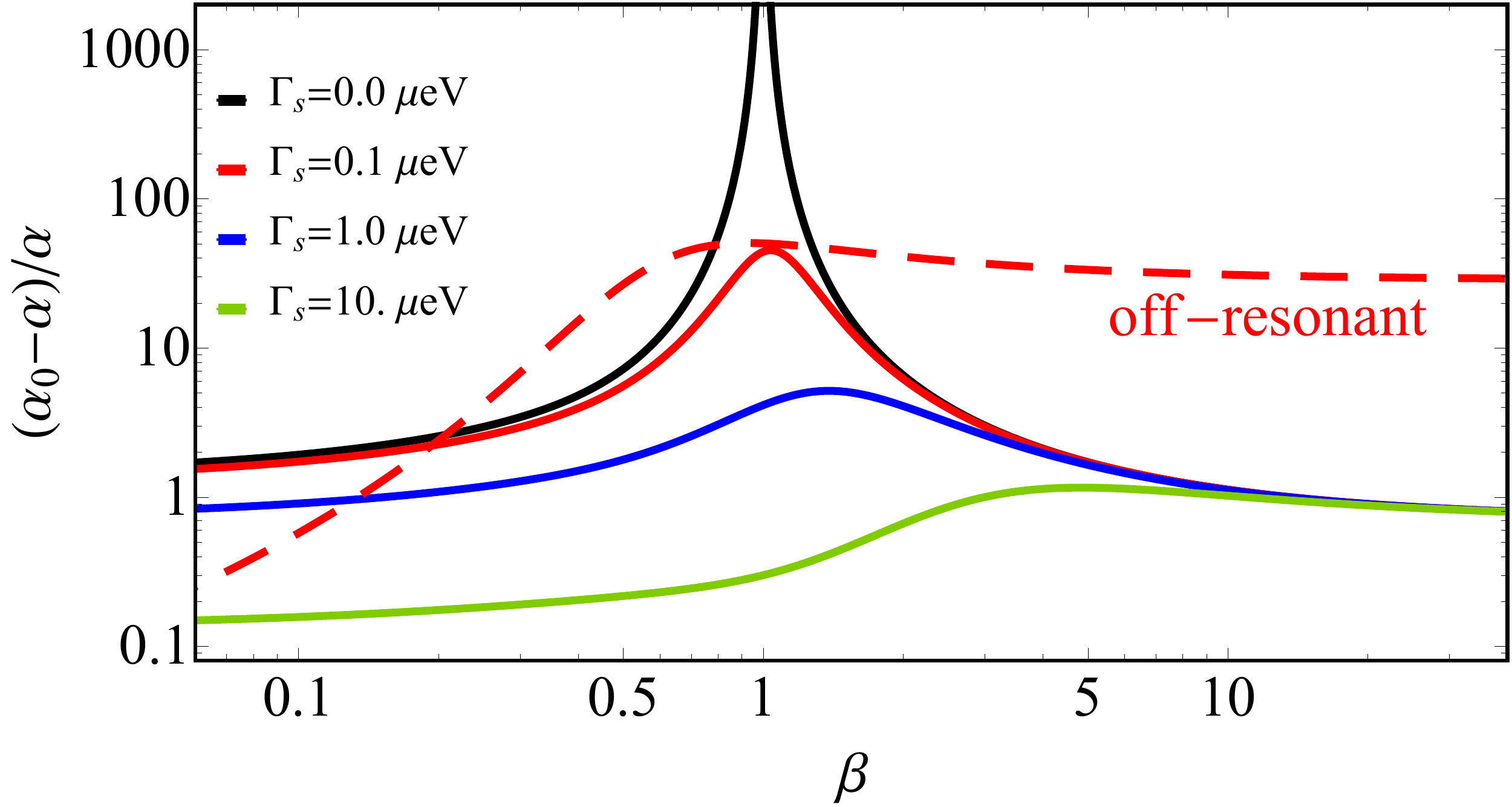}
\caption{Absorption contrast as a function of $\beta$ for different indicated values of $\Gamma_s$ on $2p$-exciton resonance and two-photon resonance ($\delta_p=\delta_s=0$). $\beta$ is varied by changing $\Omega$ while $\Omega_s=50~\mu$eV is held constant, and the remaining parameters coincide with those of Fig.~\ref{fig.EIT_ps}. The dashed line shows the absorption contrast for $\Gamma_s = 0.1~\mu$eV and probe laser tuned away from the $2p$ resonance at $\delta_p=20$~meV. $\beta$ is varied by changing $\Omega_s$ with $\Omega = 50~\mu$eV.}\label{fig.EITdelS}
\end{figure}

A key ingredient to leverage the power of Rydberg interactions optically is a large interaction-induced absorption contrast, which would result in a large nonlinearity due to a blockade of multiple Rydberg excitations \cite{pritchard2010}. Hence, the absorption contrast, $(\alpha_0-\alpha)/\alpha$, between single-photon driving, $\alpha_0 = \alpha_\text{res} + \alpha_\text{bg}$, and two-photon excitation, $\alpha$, provides a suitable figure of merit for estimating the extent of blockade effects on the optical response. As shown in Fig.~\ref{fig.EITdelS}, the contrast is maximal for $\Gamma_s=0$ and around $\beta=1$. Since the non-dissipative dark state established under EIT conditions is perturbed by decoherence and decay of the Rydberg state, the contrast tends to decrease as the decay rate $\Gamma_s$ increases. While precise values for $\Gamma_s$ remain to be determined experimentally, one can expect comparable coherence properties as for $p$-state Rydberg excitons, suggesting that $\Gamma_s$ is in the $\mu$eV range for the highest principal quantum numbers~\cite{RydExc}. This suggests that considerable blockade-induced absorption contrasts of $(\alpha_0-\alpha)/\alpha \approx 3$ are feasible.

While resonant $p$-state coupling maximizes the achievable contrast for optimal values of $\beta$, it also implies a strong $\beta$-dependence that originates from the interference between the two different excitation pathways, as discussed above. An interesting  alternative approach is, therefore, to deliberately operate far off any $p$-state resonance and establish conditions of EIT only via the background. Consequently, this offers a robust way to established dark exciton states in the material. As demonstrated in Fig.~\ref{fig.system}(c), the realization of such phonon-mediated EIT conditions indeed yields a spectrum of sharp transmission resonances, where the absorption drops to very small values whenever one hits a two-photon resonance with an excited Rydberg $s$-state. EIT in this case is indeed enabled predominantly by the continuum of phonon-dressed exciton states, since the probe laser is tuned right between the $2p$ and $3p$ resonances to render $p$-state excitation inefficient. As a result, the absorption drops to almost zero at each transmission resonance, limited only by the Rydberg state linewidth, assumed $\Gamma_s=5$~neV for the $1s$-para exciton and $0.8/n^3$~meV for the Rydberg states with $n>1$.

As illustrated in Fig.~\ref{fig.EITdelS}, this phonon-mediated EIT also renders the absorption contrast much more insensitive to variations of $\beta$, since the Rydberg blockade can now switch solely between background-mediated EIT transmission and strong absorption via phonon-assisted exciton coupling. Thereby achievable absolute values of the absorption contrast are around $\alpha_0-\alpha \sim 30$~mm$^{-1}$. For a typical blockade radius of about 5~$\mu$m~\cite{RydExc}, this corresponds to optical depths per blockade volume~\cite{Gorshkov2011a} of $OD_b \approx 0.3$. These values are comparable to those obtained in magneto-optical traps for atomic gases~\cite{pritchard2010} and would therefore already provide a platform for observing highly nonlinear light propagation with low-intensity coherent fields~\cite{pritchard2010,sevincli2011,walther2018giant}. 

In conclusion, we have presented a microscopic description for electromagnetically induced transparency with excitons in a semiconductor. Focusing on the yellow exciton series in Cu$_2$O and its associated phonon-assisted background, we have shown that optical coupling to a long-lived $s$-state can be used to control the effects of such phonon-induced processes via the formation of an EIT dark state. The resulting EIT absorption offers a wide range of exciting spectroscopic possibilities to suppress and separate the excitonic and phononic contributions, and thereby opens up new experimental capabilities to deepen our understanding of the optical properties of semiconductors. We have identified a single parameter, $\beta$, that controls this behavior and can therefore be determined in future measurements.
Various implementations, based on different choices for the final $s$-state, are possible. For example, by coupling to the $1s$-para exciton of the yellow series in Cu$_2$O, one could benefit from the extraordinarily small linewidth $\Gamma_s \sim 5$ neV of this state~\cite{Koirala2013}. The narrow EIT resonances possible in such a scenario may be exploited to realize enhanced Kerr nonlinearities \cite{Schmidt96} to generate interactions between dark-state polaritons. 
Strong exciton interactions could be used directly by exciting high-lying Rydberg $s$-states based on the ladder scheme illustrated in Fig.~\ref{fig.system}. Here, the lower transition still optimizes the light-matter coupling and the upper transition can be driven with THz-radiation~\cite{THzLaser}. While we have focused here on Cu$_2$O semiconductors, the developed theory applies to a range of other materials, such as ZnO \cite{Ozgur2005}, where different phonon channels and energy scales may consequently yield different control capabilities of the exciton coupling and phonon-assisted absorption. Finally, the possible integration into optical cavities presents another exciting outlook. Already a moderate cavity enhancement of the probe-field coupling by a factor of $\sim10$ would make optical depths of $OD_b \gtrsim 1$ possible and may, thereby, permit generating and manipulating correlated quantum states of light \cite{Chang2014}.

\section*{Acknowledgments}
The authors thank Richard Schmidt and Marc A{\ss}mann for valuable discussions. This work has been supported by the EU through the H2020-FETOPEN Grant No. 800942640378 (ErBeStA), by the DFG through the SPP1929, by the Carlsberg Foundation through the Semper Ardens Research Project QCooL, by the DNRF through a Niels Bohr Professorship to T. P. and by the NSF through a grant for the Institute for Theoretical Atomic, Molecular, and Optical Physics at Harvard University and the Smithsonian Astrophysical Observatory.

\bibliographystyle{apsrev4-1}
%

\clearpage
\onecolumngrid
\appendix

\section{Dressed State description} \label{app:dressed_states}
In this section, we derive the excitonic equations of motion as given in Eqs.~(\ref{mt1}-\ref{mt3}) of the main text. The diagonal part of the Hamiltonian in the rotating frame is given by
\begin{equation}
	\begin{split}
		\hat H_0= \ &\hbar \int{\rm d}^3q \left(\delta_p(\mathbf q)\hat X^\dagger_p(\mathbf q)\hat X_p(\mathbf q)+\delta_s(\mathbf q)\hat X^\dagger_s(\mathbf q)\hat X_s(\mathbf q)\right)+
		\sum\limits_\mu\hbar\int{\rm d}^3q\:\delta_\mu(\mathbf q)\hat X^\dagger_\mu(\mathbf q)\hat X_\mu(\mathbf q)\\
	&+\hbar\int{\rm d}^3q\left(\delta_0(\mathbf q)\hat X_{1s}^\dagger(\mathbf q)\hat X_{1s}(\mathbf q)+\omega(\mathbf q)\hat b^\dagger(\mathbf q)\hat b(\mathbf q)\right),
	\end{split}\label{eq.Hdiag}
\end{equation}
where the detunings of the yellow $p$-exciton, $\hbar\delta_{p}({\mathbf q})=\varepsilon_{p}({\mathbf q})-\hbar\omega_p$, and the blue s-exciton, $\hbar\delta_{\mu}({\mathbf q})=\varepsilon_{\mu}({\mathbf q})-\hbar\omega_p$, are given in terms of the respective exciton energies $E_p$ and $E_\mu$, and the frequency $\omega_p$ of the probe laser. The energies $\varepsilon_p=E_p+\hbar^2q^2/(2m)$ and $\varepsilon_\mu=E_\mu+\hbar^2q^2/(2m)$ are composed of the respective excitation energy of the exciton state and the kinetic energy $\hbar^2q^2/(2m)$ of the excitons. Similarly, the two-photon detuning of the yellow Rydberg $s$-exciton, $\hbar\delta_s({\mathbf q})=\varepsilon_s({\mathbf q})-\hbar\omega_p-\hbar\omega_c$, is determined by the exciton energy $\varepsilon_s=E_s+\hbar^2q^2/(2m)$ and the combined frequency of the probe ($\omega_p$) and control ($\omega_c$) laser beam. 

The operators $\hat X_{p}(\mathbf q)$, $\hat X_{s}(\mathbf q)$ and $\hat X_{\mu}(\mathbf q)$ are bosonic annihilation operators of the yellow $p$-exciton, the yellow Rydberg $s$-exciton and blue $s$-excitons, respectively. Finally, the operator $\hat{X}_{1s}^\dagger(\mathbf q)$ creates the long-lived yellow 1$s$-ortho exciton that together with the $\Gamma_3^-$ phonon branch, described by annihilation operators $\hat b(\mathbf q)$, is responsible for the background absorption. Their respective energies are given by $\varepsilon_0=E_0+\hbar^2q^2/(2m)$, giving rise to $\hbar\delta_0(\mathbf q) = \varepsilon_0 - \hbar \omega_{\rm p}$, and $\omega(\mathbf q)=\omega_{\Gamma_{3}^-}={\rm const.}$, which does not depend on the momentum of the optical phonons.
The exciton-phonon coupling that leads to the generation of the broad absorption background can be well described by a Fr\"ohlich interaction term
\begin{equation}
	\begin{split}
		\hat H_F=&\sum\limits_{\mu}\int {\rm d}^3q\:{\rm d}^3q^{\prime}\,h_{\mu}(\mathbf q,\mathbf q^{\prime})[\hat b^\dagger(-\mathbf q)\hat X_{1s}^\dagger(\mathbf q+\mathbf q^{\prime})\hat X_\mu(\mathbf q^{\prime})+\text{h.c.}].
	\end{split}
\end{equation}
To shorten notation, we define a composite Boson operator of the combined state of a yellow $1s$-exciton and a phonon
\begin{equation}
	\hat X(\mathbf q,\mathbf p)=\hat b(-\mathbf q)\hat{X}_{1s}(\mathbf q+\mathbf p).
\end{equation}
The light-matter coupling is described by the Hamiltonian 
\begin{equation}
	\begin{split}
		\hat H_L= \ &\hbar g_p\int{\rm d}^3q\:\mathcal E(\mathbf q)\left(\hat X_p(\mathbf q)+\hat X_p^\dagger(\mathbf q)\right)
		+\sum\limits_\mu \hbar g_\mu\int{\rm d}^3q\:\mathcal E(\mathbf q)\left(\hat X_\mu(\mathbf q)+\hat X_\mu^\dagger(\mathbf q)\right)\\
		&+\hbar\Omega_{s}\int{\rm d}^3q\:\left(\hat X^\dagger_p(\mathbf q)\hat X_s(\mathbf q)+h.c.\right)
		+\hbar\Omega_{\mu}\int{\rm d}^3q\:\left(\hat X^\dagger_\mu(\mathbf q)\hat X_s(\mathbf q)+h.c.\right),
	\end{split}
\end{equation} 
where the constants $g_p$ and $g_\mu$ denote the coupling of the yellow $p$-exciton ($g_p$) and blue $s$-exciton ($g_{\mu}$) to a probe laser field with amplitude $\mathcal E(\mathbf q)$. The last two lines describe the Rydberg-exciton excitation from these intermediate-state excitons by the control laser with respective Rabi frequencies $\Omega_s$ and $\Omega_\mu$. The finite exciton lifetime and other line broadening effects are included in the description by homogeneous linewidths $\Gamma_p$, $\Gamma_s$, $\Gamma_\mu$, and $\Gamma$ for the yellow $p$-exciton, the yellow $s$-exciton, the blue $s$-excitons, and the phonon-dressed $1s$-exciton, respectively. The corresponding Heisenberg equations of motion for the field operators can then be written as
\begin{align}
 \dot{\hat{X}}_p({\mathbf q})=&-\gamma_p \hat{X}_p({\mathbf q})-ig_p\mathcal E({\mathbf q})-i\Omega_s \hat{X}_s({\mathbf q})\\
 \dot{\hat{X}}_\mu({\mathbf q})=&-\gamma_\mu \hat{X}_\mu({\mathbf q})-ig_\mu\mathcal E({\mathbf q})-i\Omega_\mu \hat{X}_s({\mathbf q})\label{eq6}-i\int {\rm d}^3  q^\prime\:h_\mu(\mathbf q^\prime, \mathbf q) \hat{X}(\mathbf q^\prime,\mathbf q) \\
 \dot{\hat{X}}_s({\mathbf q})=&-\gamma_s \hat{X}_s({\mathbf q})-i\Omega_s \hat{X}_p({\mathbf q})-i\sum\limits_\mu \Omega_\mu \hat{X}_\mu({\mathbf q}) \\
 \dot{\hat{X}}(\mathbf q,\mathbf q^\prime)=&-\gamma(\mathbf q, \mathbf q^\prime) \hat{X}(\mathbf q,\mathbf q^\prime)-i \sum\limits_\mu h_\mu(\mathbf q,\mathbf q^\prime)\hat{X}_\mu (\mathbf q^\prime)
\end{align}
Here, we have introduced the complex widths $\gamma_p=\Gamma_p+i\delta_p({\mathbf q})$, $\gamma_\mu=\Gamma_\mu+i\delta_\mu({\mathbf q})$, $\gamma_s=\Gamma_s+i\delta_s({\mathbf q})$ and $\gamma=\Gamma+i\delta({\mathbf q},\mathbf q')$. The detuning of the composite boson operator is $\delta({\mathbf q},\mathbf q') = \delta_0(\mathbf q+\mathbf q') + \omega_{\Gamma_{3}^-}$.
With an approximate value of $|\delta_\mu|\approx 0.4$~eV, the detuning of the blue excitons is much larger than any other relevant energy scale and, therefore, permits an adiabatic elimination of their dynamics, which yields from Eq.~(\ref{eq6})
\begin{align}
\hat{X}_\mu({\mathbf q})=&-i\frac{g_\mu}{\gamma_\mu}\mathcal E({\mathbf q})-i\frac{\Omega_\mu}{\gamma_\mu} \hat{X}_s({\mathbf q})-i\int {\rm d}^3 q^\prime\: \frac{h_\mu(\mathbf q^\prime, \mathbf q)}{\gamma_\mu} \hat{X}(\mathbf q^\prime,\mathbf q)\label{Xmu}.
\end{align}
Substituting this result into the remaining equations of motion yields the following coupled dynamics 
\begin{align}
 \dot{\hat{X}}_s({\mathbf q})=&-\gamma_s \hat{X}_s({\mathbf q})-i\Omega_s \hat{X}_p({\mathbf q})+ig_b\mathcal{E}({\mathbf q})+i\Omega_b\hat{X}_s({\mathbf q})\label{gb}+i\int {\rm d}^3 q^\prime\Omega(\mathbf q^\prime,\mathbf q) \hat{X}(\mathbf q^\prime,\mathbf q) \\
 \dot{\hat{X}}(\mathbf q,\mathbf q^\prime)=&-\gamma(\mathbf q, \mathbf q^\prime) \hat{X}(\mathbf q,\mathbf q^\prime)+ig(\mathbf q,\mathbf q^\prime)\mathcal E({\mathbf q}^\prime)
+i \Omega(\mathbf q,\mathbf q^\prime) \hat{X}_s({\mathbf q}^\prime)
 +i\!\int {\rm d}^3 q^{\prime\prime}C(\mathbf q,\mathbf q^\prime,\mathbf q^{\prime\prime}) \hat{X}(\mathbf q^{\prime\prime},\mathbf q^\prime)\label{excphon}
\end{align}
with the effective coupling strengths
\begin{align}
	g_b=&\sum\limits_\mu\frac{g_\mu\Omega_\mu}{\delta_\mu},\ \Omega_b=\sum\limits_\mu\frac{\Omega_\mu^2}{\delta_\mu},\\
	g(\mathbf q,\mathbf q^\prime)=&\sum\limits_\mu g_\mu\frac{h_\mu(\mathbf q,\mathbf q^\prime)}{\delta_\mu},\\
	\Omega(\mathbf q,\mathbf q^\prime)=&\sum\limits_\mu \Omega_\mu\frac{h_\mu(\mathbf q,\mathbf q^\prime)}{\delta_\mu}\\	
	C(\mathbf q,\mathbf q^\prime,\mathbf q^{\prime\prime})=&\sum\limits_\mu\frac{h_\mu(\mathbf q,\mathbf q^\prime)h_\mu(\mathbf q^{\prime\prime},\mathbf q^\prime)}{\delta_\mu}.
\end{align}
to leading order in $\Gamma_\mu/|\delta_\mu|\ll1$.

Let us now discuss the relative contributions from the various excitation processes and interactions. While the optical coupling strengths $g_\mu$ and $\Omega_\mu$ can be on the same order as $g_p$ and $\Omega_p$, the detuning $|\delta_\mu|$ is much larger than $g_\mu$ and $\Omega_\mu$. This implies that the $g_b$ is typically much smaller than $g_p$, such that we can neglect its direct contribution to the probe-light absorption and discard the corresponding coupling term in Eq.~(\ref{gb}). Likewise, $\Omega_b$ is typically very small and therefore only causes an insignificant shift of the Rydberg exciton line [cf. Eq.~(\ref{gb})] and can safely be neglected compared to its width $\Gamma_s$. The last term in Eq.~(\ref{excphon}) describes a change of phonon momenta via virtual exciton-phonon interactions. This second-order process is strongly suppressed by the large energy offset $\delta_{\mu}$ and, therefore, can also be neglected. Note, however, that these terms are discarded merely to simplify the resulting expressions and can otherwise be straightforwardly included in the calculations. With these additional simplifications, we obtain Eqs.~(\ref{mt1}-\ref{mt3})
\begin{align}
 \dot{\hat{X}}_p({\mathbf q})=&-\gamma_p \hat{X}_p({\mathbf q})-ig_p\mathcal E({\mathbf q})-i\Omega_s \hat{X}_s({\mathbf q})\label{mt1_app}\\
 \dot{\hat{X}}_s({\mathbf q})=&-\gamma_s \hat{X}_s({\mathbf q})-i\Omega_s \hat{X}_p({\mathbf q}) +i\int {\rm d}^3 q^\prime\Omega(\mathbf q^\prime,\mathbf q) \hat{X}(\mathbf q^\prime,\mathbf q) \label{mt2_app}\\
 \dot{\hat{X}}(\mathbf q,\mathbf q^\prime)=&-\gamma(\mathbf q, \mathbf q^\prime) \hat{X}(\mathbf q,\mathbf q^\prime)+ig(\mathbf q,\mathbf q^\prime)\mathcal E({\mathbf q}^\prime)
 +i \Omega(\mathbf q,\mathbf q^\prime) \hat{X}_s({\mathbf q}^\prime)\label{mt3_app}
\end{align}
given in the main text.

\section{Calculating the optical response} \label{app:opt_res}
The propagation of the probe field amplitude through the sample is determined by
\begin{equation}
\partial_t\mathcal E(\mathbf q)+\frac{c}{n}\partial_z\mathcal E(\mathbf q)=-i\frac{g_p}{n^2}\langle\hat X_p(\mathbf q)\rangle-i\sum_\mu \frac{g_\mu}{n^2}\langle\hat X_\mu(\mathbf q)\rangle
\end{equation}
where $n=2.74$ denotes the dielectric constant of Cu$_2$O. The susceptibility is thus obtained from
\begin{equation}
i\chi^{(1)}(\mathbf q)\mathcal{E}(\mathbf q) =-i\frac{g_p}{cn}\langle\hat X_p(\mathbf q)\rangle-i\sum_\mu \frac{g_\mu}{cn}\langle\hat X_\mu(\mathbf q)\rangle
\end{equation}
and its imaginary part determines the inverse absorption length $\alpha=\Im(\chi^{(1)})$. Substituting Eq.~(\ref{Xmu}) and again neglecting terms proportional to $\Omega_\mu/\gamma_\mu$ and $g_\mu/\gamma_\mu$, this can be written as
\begin{equation}
	\begin{split}
		i\chi^{(1)}(\mathbf q)\mathcal{E}(\mathbf q) =&-i\frac{g_p}{cn}\langle\hat X_p (\mathbf q)\rangle+i\int {\rm d}^3q^\prime\: \frac{g(\mathbf q^\prime, \mathbf q)}{cn} \langle\hat{X}(\mathbf q^\prime,\mathbf q)\rangle
	\end{split}
\end{equation}
in terms of the phonon-dressed $1s$-exciton amplitude.
Its steady-state value
\begin{align}
 \langle \hat{X}(\mathbf q,\mathbf q^\prime) \rangle = i\frac{g(\mathbf q,\mathbf q^\prime)}{\gamma(\mathbf q, \mathbf q^\prime)}\mathcal E({\mathbf q}^\prime)+i \frac{\Omega(\mathbf q,\mathbf q^\prime)}{\gamma(\mathbf q, \mathbf q^\prime)} \langle \hat{X}_s({\mathbf q}^\prime) \rangle.\label{Xqq}
\end{align}
is obtained by taking expectation values over the equations of motion Eqs.~(\ref{mt1_app}-\ref{mt3_app}) and solving for the steady state of Eq.~(\ref{mt3_app}).
Using this result in Eq.~(\ref{mt2}) yields for the steady-state amplitude of the Rydberg exciton 
\begin{align}
\langle \hat{X}_s({\mathbf q}) \rangle =-i\frac{\Omega_s}{{\gamma_s+D_{\Omega^2}}} \langle \hat{X}_p({\mathbf q}) \rangle -\frac{D_{g\Omega}}{{\gamma_s+D_{\Omega^2}}}\mathcal E({\mathbf q})\label{Xs}
\end{align}
where  
\begin{align}
D_{g\Omega}=&\int {\rm d}^3 q^\prime \frac{\Omega(\mathbf q^\prime,\mathbf q)g(\mathbf q^\prime,\mathbf q)}{\gamma(\mathbf q^\prime, \mathbf q)},\label{eq.DgOint_app}\\
D_{\Omega^2}=& \int {\rm d}^3 q^\prime\frac{\Omega(\mathbf q^\prime,\mathbf q)^2}{\gamma(\mathbf q^\prime, \mathbf q)},\label{eq.DO2int_app}\\
D_{g^2}=& \int {\rm d}^3 q^\prime\frac{g(\mathbf q^\prime,\mathbf q)^2}{\gamma(\mathbf q^\prime, \mathbf q)}.\label{eq.Dg2int_app}
\end{align}
We substitute Eq.~(\ref{Xs}) into Eq.~(\ref{mt1_app}) and Eq.~(\ref{Xqq}) to obtain
\begin{align}
\langle \hat{X}_p({\mathbf q}) \rangle =\ &i\frac{\Omega_sD_{g\Omega}-g_p({\gamma_s+D_{\Omega^2}})}{\gamma_p({\gamma_s+D_{\Omega^2}}) +\Omega_s^2}\mathcal E({\mathbf q})\\
 \langle \hat{X}(\mathbf q^\prime,\mathbf q)\rangle =\ &i\left(\frac{g(\mathbf q^\prime,\mathbf q)}{\gamma(\mathbf q^\prime, \mathbf q)}- \frac{\Omega(\mathbf q^\prime,\mathbf q)}{\gamma(\mathbf q^\prime, \mathbf q)} \frac{D_{g\Omega}}{{\gamma_s+D_{\Omega^2}}}\right)\mathcal E({\mathbf q})
+ \frac{\Omega(\mathbf q^\prime,\mathbf q)}{\gamma(\mathbf q^\prime, \mathbf q)} \frac{\Omega_s}{{\gamma_s+D_{\Omega^2}}} \langle \hat{X}_p({\mathbf q})\rangle
\end{align}
which finally gives
\begin{align}
i\chi^{(1)}(\mathbf q)=\ &\frac{g_p}{cn} \frac{\Omega_sD_{g\Omega}-g_p({\gamma_s+D_{\Omega^2}})}{\gamma_p({\gamma_s+D_{\Omega^2}}) +\Omega_s^2}
-\frac{D_{g^2}}{cn}+ \frac{D_{g\Omega}}{{cn}}\frac{D_{g\Omega}\gamma_p +g_p\Omega_s}{\gamma_p({\gamma_s+D_{\Omega^2}}) +\Omega_s^2} 
\end{align}
as given in Eq.~(\ref{eq.fullresponse}) of the main text. 

\section{Integral for background term $D_{g^2}$} \label{app:background}
\begin{figure}[ht]
\begin{center}
	\includegraphics[width=7cm]{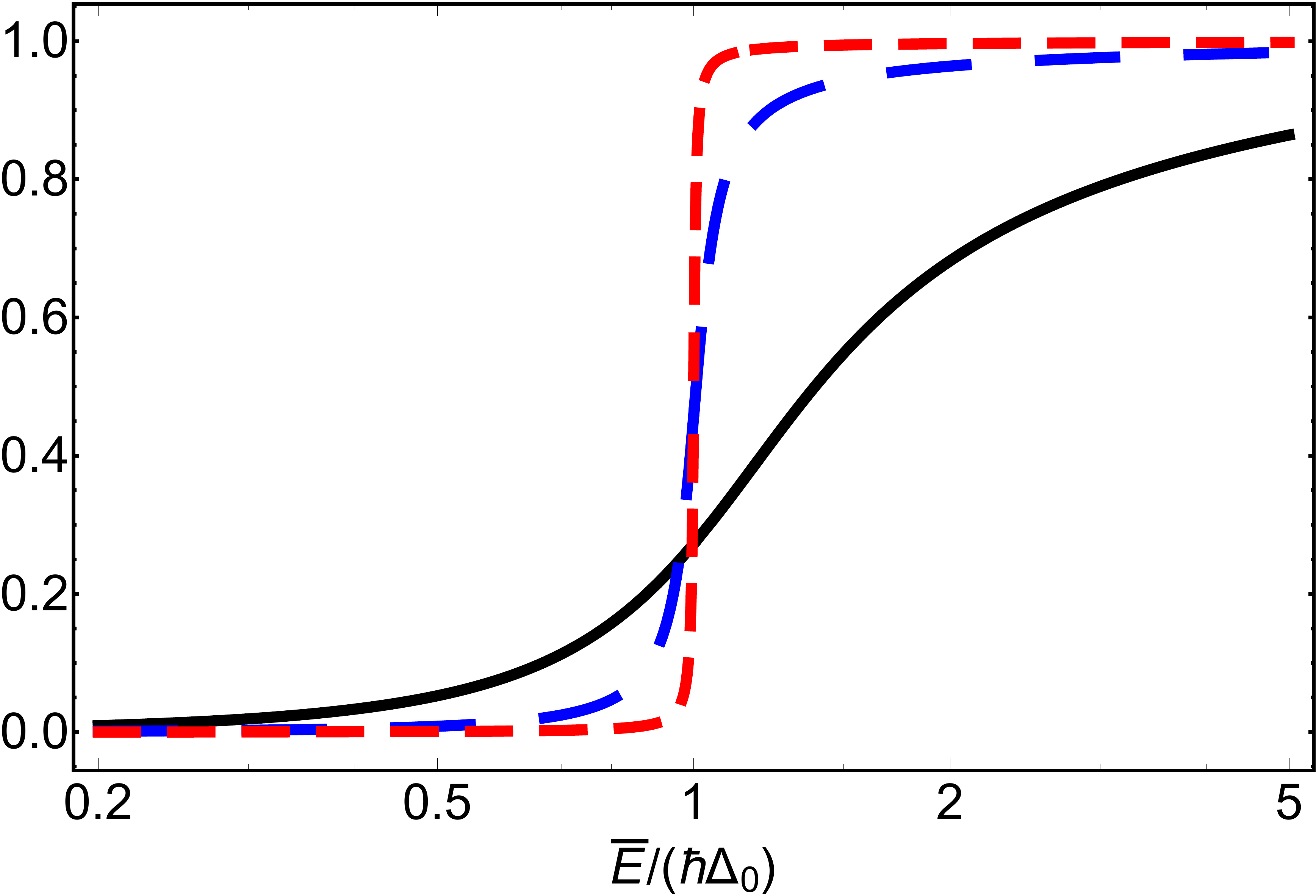}
	\includegraphics[width=7cm]{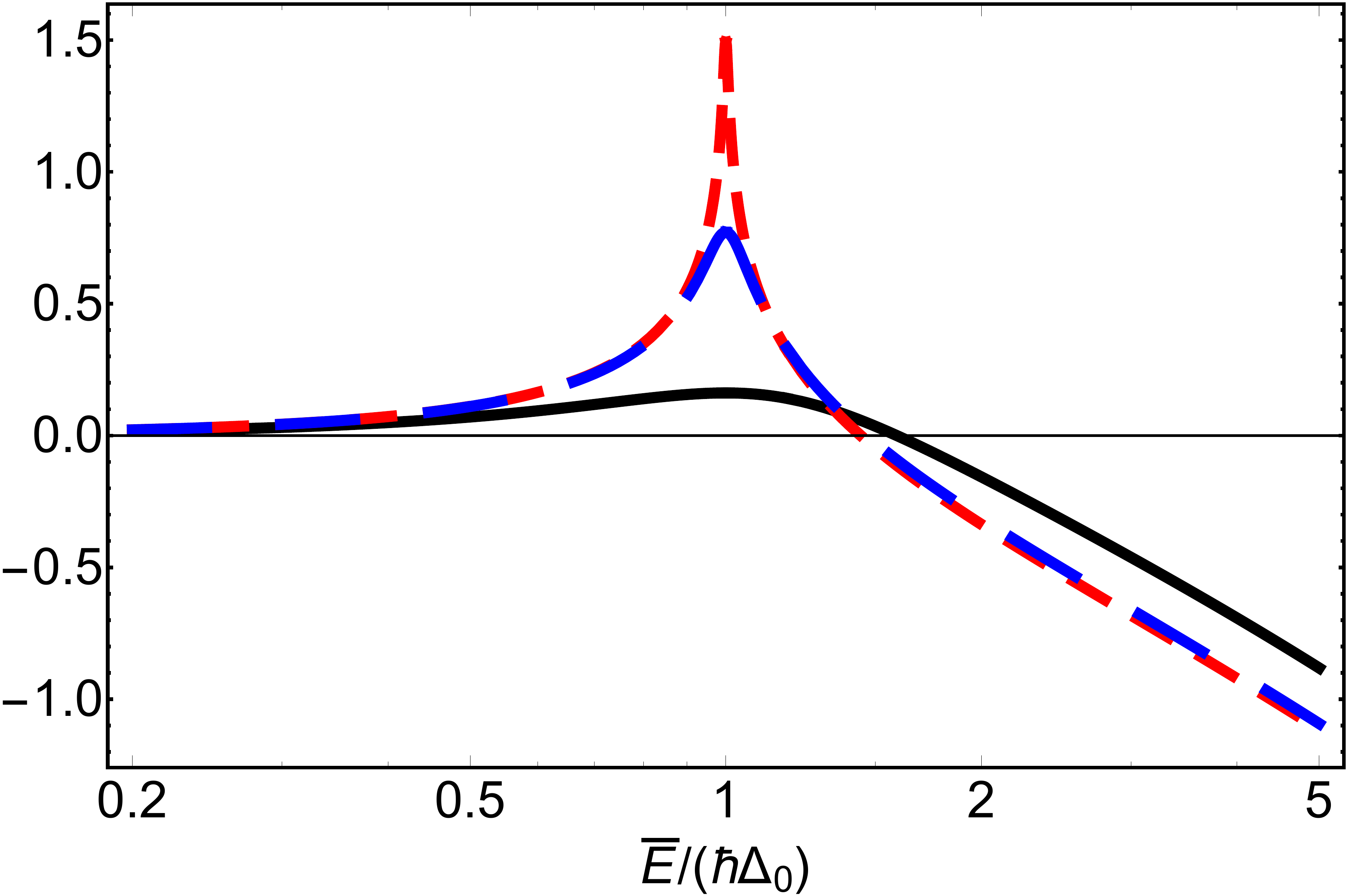}
	\caption{(Color online). Normalized integral of Eq.~(\ref{eq.intabs}) for $2\Gamma=\Delta_0, 10^{-1}\Delta_0, 10^{-2}\Delta_0$ (black, red, blue). Left: real part, right: imaginary part.}\label{fig.intabs}
\end{center}
\end{figure}

Here, we derive the solution for the integral Eq.~(\ref{eq.Dg2int_app}) given in Eq.~(\ref{eq.Dg2fin}) from the main text.
Due to the flat interaction coefficient, we have $g(\mathbf q',\mathbf q)=g$ and only the detuning remains momentum-dependent, yielding
\begin{equation}
	D_{g^2}=g^2\int {\rm d}^3q\frac{1}{\Gamma+i\delta({\mathbf q})}.
\end{equation}
Herein $\mathbf q$ is the phonon momentum, as the photon momentum is fixed by the probe laser and thus a constant.
As the phonon-assisted $1s$-excitons are metastable, the sum should be evaluated with the Dirac identity (Sokhotski-Plemelj theorem along real line integration)
\begin{equation}
	\lim\limits_{\Gamma\to 0}\frac{1}{\Gamma+i\delta({\mathbf q})}=\pi\delta_\text{Dirac}[\delta({\mathbf q})]-i\mathcal P\frac{1}{\delta({\mathbf q})},
\end{equation}
with $\mathcal P$ denoting the Cauchy principal value. The integral can also be solved for the more general case of $\Gamma>0$. For spherical symmetry the integral over the phonon momenta $\mathbf q$ simplifies as
\begin{equation}
	\int {\rm d}^3 q = 4\pi\int\limits_0^Q\,{\rm d}q\,q^2, 
\end{equation}
wherein $Q$ is the limitation of the possible phonon momenta within the crystal. For later calculations, this boundary is set at the 1st Brillouin zone $Q=\pi/a$ with the lattice constant $a$. Converting to an integral of energies $\epsilon$ and setting $\bar E=\tfrac{\hbar^2Q^2}{2m}$, the integral (without the prefactors) can be written as
\begin{align}
	\int\limits_0^{\bar E}d\epsilon\frac{\sqrt{\epsilon}}{\hbar\Gamma+i(\epsilon-\hbar\Delta_0)}=&-2i\sqrt{\bar E}\left[1+\frac{i}{z}\arctan(iz)\right]\label{eq.intabs}\\
	z=&\sqrt{\frac{\bar E}{i\hbar\Gamma+\hbar\Delta_0}},\quad \Delta_0=\omega_p-E_{0}/\hbar-\omega_{\Gamma_3^-}.
\end{align}
This gives a total structure of the term $D_{g^2}$
\begin{align}
	D_{g^2}=&-g^24\sqrt 2\pi\frac{m^{3/2}}{\hbar^2}\left\{2i\sqrt{\bar E}\left[1+\frac{i}{z}\arctan(iz)\right]\right\}.\label{eq.Dg2}
\end{align}
For large integration limits and for sufficiently low $\Gamma$ the real part of $D_{g^2}$ becomes indeed a Heaviside function with the jump at $\bar E=\hbar\Delta_0$, see Fig.~\ref{fig.intabs}.
To get back to the case of the Dirac identity, we let $\Gamma\to0$ to obtain
\begin{align}
	\lim\limits_{\Gamma\to 0}D_{g^2}=&-g^24\sqrt 2\pi\frac{m^{3/2}}{\hbar^2}\left\{2i\sqrt{\bar E}\left[1+i\sqrt{\frac{\hbar\Delta_0}{\bar E}}\arctan\left(i\sqrt{\frac{\bar E}{\hbar\Delta_0}}\right)\right]\right\}.\label{eq.Dg2fin_app}
\end{align}
As only $\bar E>\hbar\Delta_0$ is relevant for our purpose, we can use the general relation 
\begin{equation}
	\arctan(ix)=\frac{\pi}{2}\,\Theta(x-1)+i\frac{1}{2}\ln\left|\frac{1+x}{1-x}\right|,
\end{equation}
which holds for positive $x$ and thus reobtain the real part from Eq.~(\ref{eq.Dg2fin}) in the main text as
\begin{align}
	\Re\{D_{g^2}\}= 8\pi^2 g^2\frac{m^\frac{3}{2}}{\sqrt{2}\hbar^2}\sqrt{\hbar\Delta_0}.
\end{align}
This result is identical to the direct solution with the Dirac identity. For the imaginary part we obtain
\begin{align}
	\Im\{D_{g^2}\}=&-8 \pi g^2\frac{m^\frac{3}{2}}{\sqrt{2}\hbar^2}\left\{2\sqrt{\bar E}\left[1-\sqrt{\frac{\hbar\Delta_0}{4\bar E}}\ln\left|\frac{\sqrt{\bar E}+\sqrt{\hbar\Delta_0}}{\sqrt{\bar E}-\sqrt{\hbar\Delta_0}}\right|\right]\right\}.
\end{align}
In case $\bar E\gg \hbar\Delta_0$, the last term in square brackets will become negligible, yielding a constant imaginary part of
\begin{align}
	\Im\{D_{g^2}\}=-16\pi g^2\frac{m^\frac{3}{2}}{\sqrt{2}\hbar^2}\sqrt{\bar E},
\end{align}
which is exactly the imaginary part of Eq.~(\ref{eq.Dg2fin}) in the main text.

\section{Mathematical description of special cases of the absorption}\label{app:limiting_cases}
In this section we analyze the solution of the susceptibility in the different limiting cases depicted in Figs.~\ref{fig.EIT_ps}-\ref{fig.EITdelS} in the main text.
For the sake of brevity, we introduce an equivalent notation to Eq.~(\ref{eq.DgOint_app}-\ref{eq.Dg2int_app}) for the couplings on the exciton path,
\begin{align}
	P_{\Omega^2}=&\frac{\Omega_{s}^2}{\gamma_p},\quad P_{g\Omega}=\frac{g_p\Omega_s}{\gamma_p},\quad P_{g^2}=\frac{g_p^2}{\gamma_p}.\label{eq.Pcoeff}
\end{align}
Note that the fixed ratios of these coefficients ($P_{g\Omega}=\sqrt{P_{g^2}P_{\Omega^2}}$ etc.) are already given. Then, the susceptibility becomes
\begin{align}
 i\chi^{(1)} = &-\frac{\gamma_s(P_{g^2}+D_{g^2})+D_{g^2}P_{\Omega^2}+P_{g^2}D_{\Omega^2}-2D_{g\Omega}P_{g\Omega}}{cn(\gamma_s+D_{\Omega^2}+P_{\Omega^2})}.
\end{align}
Symmetry for exchanging the exciton and background path ($P_{XY}\leftrightarrow D_{XY}$) is obvious in this form. Yet, due to the rather different behavior of the $D$'s and $P$'s under tuning of the probe laser, asymmetric absorption spectra emerge as seen in Fig.~\ref{fig.EIT_ps} of the main text. 
Furthermore, using the relations $D_{g\Omega}=\sqrt{D_{g^2}D_{\Omega^2}}$ etc. of the main text, we can simplify the result to
\begin{align}
 i\chi^{(1)} = &-\frac{\gamma_s(P_{g^2}+D_{g^2})+(\sqrt{D_{g^2}P_{\Omega^2}}-\sqrt{P_{g^2}D_{\Omega^2}})^2}{cn(\gamma_s+D_{\Omega^2}+P_{\Omega^2})}.\label{eq.Chi1}
\end{align}
Using the quantity $\beta^2=(P_{g^2}D_{\Omega^2}/(D_{g^2}P_{\Omega^2})$, we can write this as
\begin{align}
 icn\chi^{(1)} = &-\frac{\bar\gamma_s\left(\frac{P_{g^2}}{D_{g^2}}+1\right)+(1-\beta)^2}{\frac{1}{D_{g^2}}(\bar\gamma_s+1)+\frac{\beta^2}{P_{g^2}}},\label{eq.Chi2}
\end{align}
where $\bar\gamma_s=\gamma_s/P_{\Omega^2}$ compares the Rydberg linewidth to the EIT-linewidth as a figure of merit. Hence, we can determine the full response only from the absorption in single-photon experiments, an effective Rydberg linewidth and the parameter $\beta$ relating the coupling Rabi strengths of the two paths. Congruent EIT is naturally given for $\beta\to1$ and $\bar\gamma_s\to0$. 

The inverse absorption length is given by $\alpha=\Im[\chi^{(1)}]$. Let us consider the three regimes $\beta\to0,\beta\approx 1,\beta\to\infty$ for arbitrary $\bar\gamma_s$.
For $\beta\to0$, that is, dominant exciton path, we find
\begin{align}
 icn\chi^{(1)} = -D_{g^2}-\frac{\bar\gamma_s}{\bar\gamma_s+1}P_{g^2}.
\end{align}
As shown by the red lines in Fig.~\ref{fig.EIT_ps} of the main text, the absorption reduces to the background without Rydberg-decay. For large decay $|\bar\gamma_s|\gg1$, we find little to no EIT effect, as the Rydberg line decays much faster than the EIT repopulation process, as visualized by the green line in Fig.~\ref{fig.EITdelS} of the main text.

For congruent EIT $\beta=1$, the response reduces to
\begin{align}
 icn\chi^{(1)} = &-\frac{\bar\gamma_s\left(P_{g^2}+D_{g^2}\right)}{\bar\gamma_s+\frac{1}{P_{g^2}}\left(P_{g^2}+D_{g^2}\right)}.
\end{align}
For $\bar\gamma_s\to0$ we indeed find vanishing absorption (green lines in Fig.~\ref{fig.EIT_ps} of the main text) and for $|\bar\gamma_s|\to\infty$, the EIT effect disappears. For small, but not negligible Rydberg decay, congruent EIT conditions yield
\begin{align}
 icn\chi^{(1)} = &-\bar\gamma_sP_{g^2}=-\gamma_s\frac{g_p^2}{\Omega_s^2}=-\gamma_s\frac{\Omega^2}{g^2}.
\end{align}
For increasing $\bar\gamma_s$ the maximum of contrast shifts and one can directly show that the corresponding minimal absorption is given at $\beta_\text{ex}=1+\bar\gamma_s$, where the susceptibility becomes
\begin{equation}
	icn\chi^{(1)} = -\frac{\bar\gamma_s}{1+\bar\gamma_s}P_{g^2}.
\end{equation}
This shift is clearly visible in Fig.~\ref{fig.EITdelS} of the main text.

Finally, for a dominant background path $\beta\to\infty$, we find
\begin{align}
 icn\chi^{(1)} = &-P_{g^2}\left(1-\frac{2}{\beta}\right).
\end{align}
As shown by the blue lines in Fig.~\ref{fig.EIT_ps} of the main text, now the background is suppressed and only the exciton line is visible in absorption. Note that, even to first-order corrections of $\beta$, there is no dependence on the Rydberg-linewidth, and only a root dependence on the EIT linewidth, which is given by $P_{\Omega^2}$. Hence, as can be seen in Fig.~\ref{fig.EITdelS} of the main text, all lines show the same large-$\beta$ limit, independent of $\Gamma_s$.

\end{document}